\documentclass[10pt,aps,prc,showpacs,nofootinbib,twocolumn]{revtex4-1}

\usepackage{epsfig}
\usepackage{graphicx,amsmath}
\usepackage{color}
\newcommand\ba{\begin{eqnarray}}
\newcommand\ea{\end{eqnarray}}

\newcommand{\be}{\begin{equation}}
\newcommand{\ee}{\end{equation}}
\newcommand{\bas}{\begin{eqnarray*}}
\newcommand{\eas}{\end{eqnarray*}}
\begin{document}
\title{\bf \large Fractal properties in fundamental force coupling constants,\\
 in atomic energies, and in elementary particle masses}

\author{B. Tatischeff\\Institut de Physique Nucléaire (UMR 8608), CNRS/IN2P3 - Univ Paris-Sud\\ F-91406  ORSAY CEDEX}

\pacs{14.20.-c, 14.40.-n, 14.60.-z, 14.65.-q, 14.70.-e}

\vspace*{1cm}
\begin{abstract}Using the discrete-scale invariance theory, we show that the coupling constants of fundamental forces, the atomic masses and energies, and the elementary particle masses, obey to the fractal properties.
\end{abstract}
\maketitle
\section{Introduction}
The large and increasing number of mesons and baryons, suggests the need for a new classification, in addition to those already existing, based on their quark and gluon nature and quantum numbers (isospin, spin, charge conjugation and parity). A possible way is to look for eventuel fractal properties of these particles. Up to now, the very powerful concept of fractals \cite{mandelbrot}, has not been really used to study the particle physics masses. Nottale, after the "Study of the Theory of Scale Relativity" which aims to unify Quantum Physics and Relativity theory \cite{nottale1} discussed several applications. He noted that the "{\it lepton e, $\mu$, and $\tau$ mass ratios followed a power-low sequence}", and that the quark mass ratio $m_{s}/m_{d}$ is close to the $e^{3}$ = 20.086 value.

In the same mind, several relations were given, relying between themselves the masses of the two quark families: $m_u, m_c$, and $m_t$ in the one hand, and 
$m_d, m_s$, and $m_b$ in the other hand \cite{btib}. The same study presented relations between gauge boson masses, and another relation between lepton masses \cite{btib}. These relations suggest that the particle physics masses should also follow fractal properties.

Before searching for possible fractal properties inside the very large field of meson and baryon masses, it is useful to study them in more fundamental particles.
The subject  of the present paper is  to look - in spite of small statistics - at possible fractal properties of fundamental coupling constant forces, at possible fractal properties of several electronic energies of atomic nuclei, and also at possible fractal properties in elementary particle masses. 

\vspace{-5.mm}
\section{Remind of the used scale invariance models}
\subsection{Fractal characteristic of a set of data}
The fractal concept states that the same physical laws apply for different scales of the given physics. We summarize here briefly the concept of continuous and discrete scale invariances, transcribing the developpements of D. Sornette \cite{sornette} and L. Nottale \cite{chaline}.
The concept of {\it continuous} scale invariance is defined in the following way: an observable O(x), function of the variable x,  is scale invariant under the arbitrary change x $\to~\lambda$x, if there is a number $\mu(\lambda)$ such that O(x) = $\mu$O($\lambda$x). $\lambda$ is the fundamental scaling ratio. The solution of  O(x) is:
\be
 O(x) = C x ^{\alpha}
\ee
where  $\alpha$ = -ln$\mu$/ln$\lambda$.

The relative value of the observable, at two different scales, depends only on $\mu$, the ratio of the two scales O(x)/O($\lambda$x) and does not depend on x. We have therefore "a continuous translational invariance expressed on the logarithms of the variables".
\subsection{Characteristics of the discrete scale invariance }
Unlike the {\it continuous} scale invariance, the {\it discrete} scale invariance (DSI) applies, when the scale invariance is only observed for specific choices of $\lambda$ \cite{sornette}
\cite{chaline}. 

 The signature of DSI is the presence of power laws with complex exponents $\alpha$ inducing log-periodic corrections to scaling. In case of DSI, the $\alpha$ exponent is now 
\be
 \alpha = -ln\mu/ln\lambda + i 2\pi~n/ln\lambda 
\ee
where n is an arbitrary integer. The {\it continuous} scale invariance is obtained for the special case n = 0, then $\alpha$ becomes real.
\section{Applications}
We apply this concept to the study of elementary physical species. The continuously varying variable "x" is now replaced by the rank: (r) of the discrete values of the studied quantity (q). First we look for possible linear fits in the log scale plot of the successive values of the studied species: $log(q_{r}) = log(r)$ which signs the fractal property. 

Then, keeping the first term of equation (2) (n~=~1), we get the most general form f(r)~=~$q_{r+1}/q_{r}$ of DSI distributions \cite{sornette}:
\be
 f(r) = C\hspace*{1.mm}(|r-r_{c}|)^{l}\hspace*{1.mm}[1 + a_{1}\hspace*{1.mm}cos(2\pi\hspace*{1.mm}\Omega\hspace*{1.mm}
ln\hspace*{1.mm}(|r-r_{c}|) + \Psi)]\\
\ee
where we have omitted the imaginary part of f(r).

  C is a normalization constant. $a_{1}$ measures the amplitude of the log-periodic correction to continuous scaling, and $\Psi$ is a phase in the cosine. "$r_{c}$" is the  critical rank which describes the transition from one phase to another.
If "$r_{c}$"  is null, the experimental oscillations would enlarge quickly with increasing rank. When the step of the oscillations is rather stable, "$r_{c}$" is undetermined, and widely larger than the experimental "r" values.

The critical exponent "l" is defined by $\mu$ = $\lambda^{l}$. Defining $\Omega$ = 1/ln$\lambda$, we obtain $\alpha$ = -l + i 2n$\pi \Omega$. For not too small statistics, we fit the ratios q$_{r+1}/q_{r}$ by the equation (3).

In summary, the signature of scale invariances is the existence of power laws. The exponent $\alpha$ is real if we have continuous scale invariances, it is complex for discrete scale invariances, and then gives rise to log-periodic corrections.
\subsection{Application to fundamental force coupling constants}
\begin{figure}[h]
\begin{center}
\caption{Log-log distribution of the fundamental force coupling constants.}
\hspace*{-3.mm}
\scalebox{0.95}[1]{
\includegraphics[bb=22 332 520 545,clip,scale=0.5]{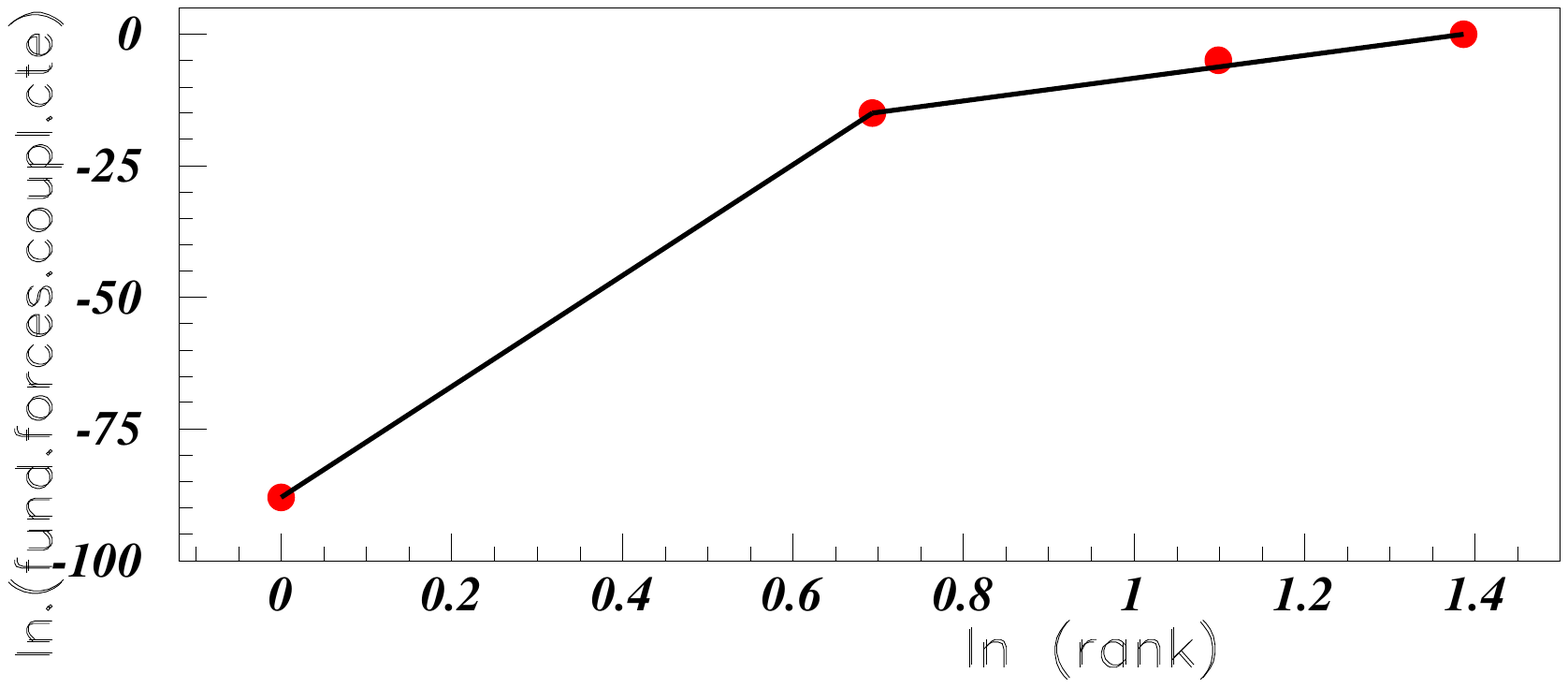}}
\end{center}
\end{figure}
The relative values of the coupling constants of the four fundamental forces, are approximately equal to: 5.9E-39 for the gravity, 3E-07 for the weak force, 1/137 for the electromagnetic force, and 1 for the strong force \cite{rohlf}. The error bars for gravity and weak force are unknown. Figure~1 shows the log-log plot of these values as a function of the log of the rank. The rank increases from the lowest force coupling constant, corresponding to gravity (rank 1)  up to the largest (strong) force coupling constant (rank 4).  The alignement between the three largest coupling constants is quite good. 
\subsection{Application to atomic masses and energies}
The Rutherford-Bohr model of the hydrogen atom, or a hydrogen-like atom(Z$\ge$1), gives  the energy levels E$_{n}$ to a rough approximation, since the Coulomb interaction between electrons is not considered.  The corresponding relation is:
\be
E_{n} = -Z^{2}R_{E}/n^{2}
\ee
A more exact calculation replaces "Z" by "Z - b" where "b" is a constant. Here 
R$_{E}$ is the Rydberg energy, "Z"  the charge and "n" the principal quantum number. The fractal property is fullfilled, since 
\be
ln (-E_{n})~=C~+~ln~(R_{E})~-~2ln (n) 
\ee
and is linear versus ln(n).
Each filled atomic shell contains 2n$^{2}$ electrons. The corresponding log-log distribution is therefore a straight line.

The photon energies emitted by a hydrogen atom, are given by the Rydberg formula :
\be
E~=R_{E}(1/n^{2}_{f}~-~1/n^{2}_{i})
\ee
the photon is emitted between the initial $n_{i}$ and final $n_{f}$ shells. 
 The log-log distribution here is not analytically aligned, but it is aligned numerically to a very good extent. This is shown in figure~2,
 where the log of theoretical values from the Rydberg formula are plotted versus the log of the rank defined by the successive  r=2$\to$1, 3$\to$2, ... transitions.
\begin{figure}[ht]
\begin{center}
\caption{Log-log distribution of photon energies given by the Rydberg formula. The log of the successive r=2$\to$1, 3$\to$2, ... transitions are plotted versus the log of the rank "r".}
\hspace*{-3.mm}
\scalebox{0.95}[1]{
\includegraphics[bb=26 230 520 544,clip,scale=0.5]{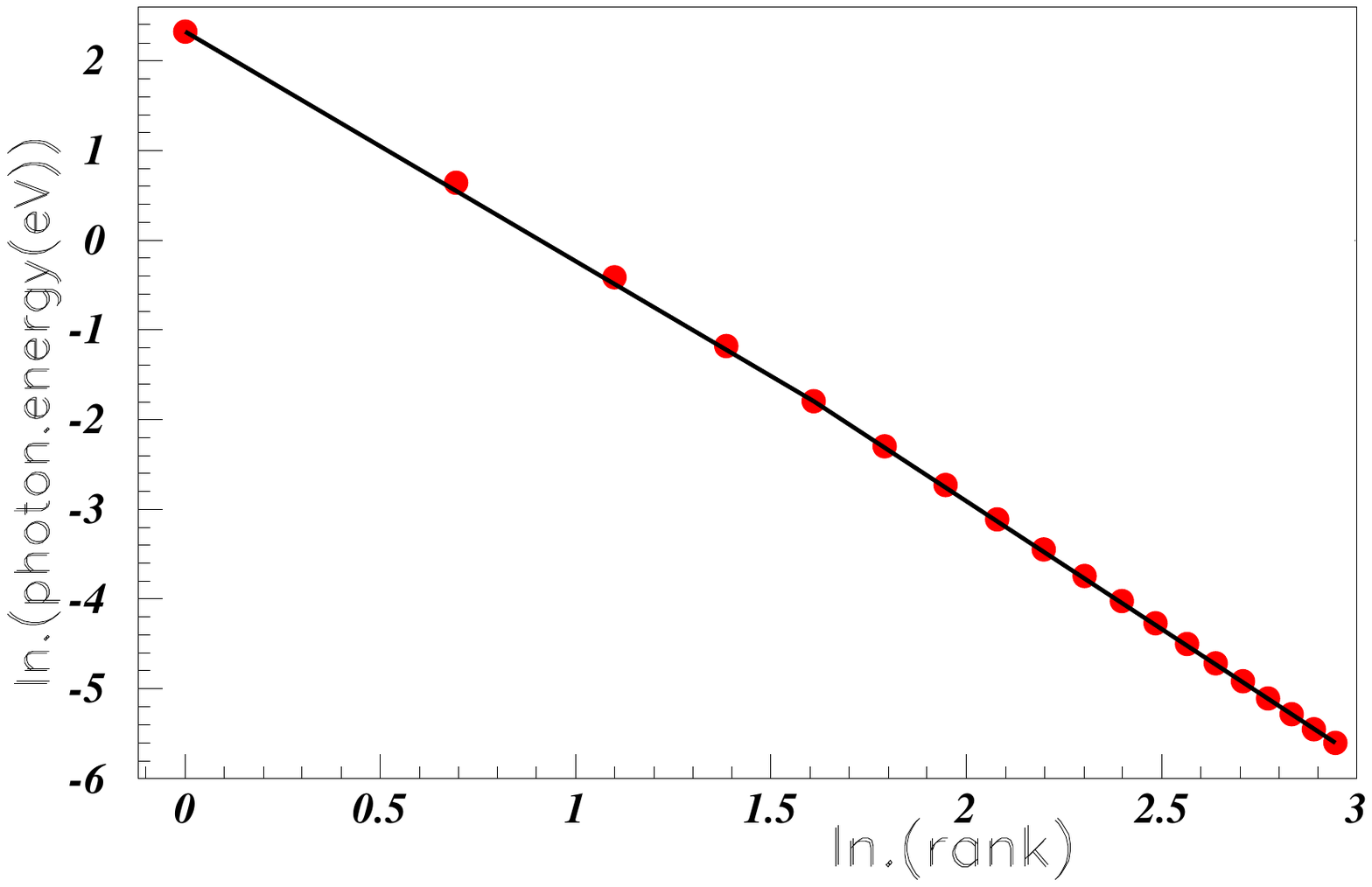}}
\end{center}
\end{figure}
\subsection{Application to elementary particle masses}
In the following, the error bars describe the experimental knowledge of the masses. The masses of the "u" and "d" quarks  are poorly defined, and  the "s" quark mass m$_{s}$=104~$\pm$~30~MeV is also unprecise; however the effect of such unprecision is negligeable in log scale. The neutrino $\nu_{\mu}$ and 
$\nu_{\tau}$ experimental masses are very badly known, since they are the
three mass-squared differences which are better known. Therefore, in addition to the experimental error bars, we plot for these particles, in the next figures, the "calculated" mass values \cite{btib} obtained using relations connecting between themselves all masses of each family.

 Figure~3 shows, in insert (a), the log-log distribution of quark masses versus the rank of the masses increasing from the lowest up to the largest value. The experimental kept values of the "u" and "d" quark masses, are those of table~I of \cite{btib}. The different slopes define the range of different fundamental scaling ratios $\lambda$.
 
  Figure~3 insert (b), shows the log-log distribution of lepton masses versus the rank of the masses increasing from the lowest up to the largest value. Here again the experimental unprecisions and "calculated"  masses of table III of \cite{btib} are used. 
\begin{figure}[h]
\begin{center}
\caption{Insert (a) shows the log-log representation of quark masses; insert (b) shows the log-log representation of lepton masses, (see text).}
\hspace*{-3.mm}
\scalebox{0.95}[1]{
\includegraphics[bb=21 130 517 558,clip,scale=0.5]{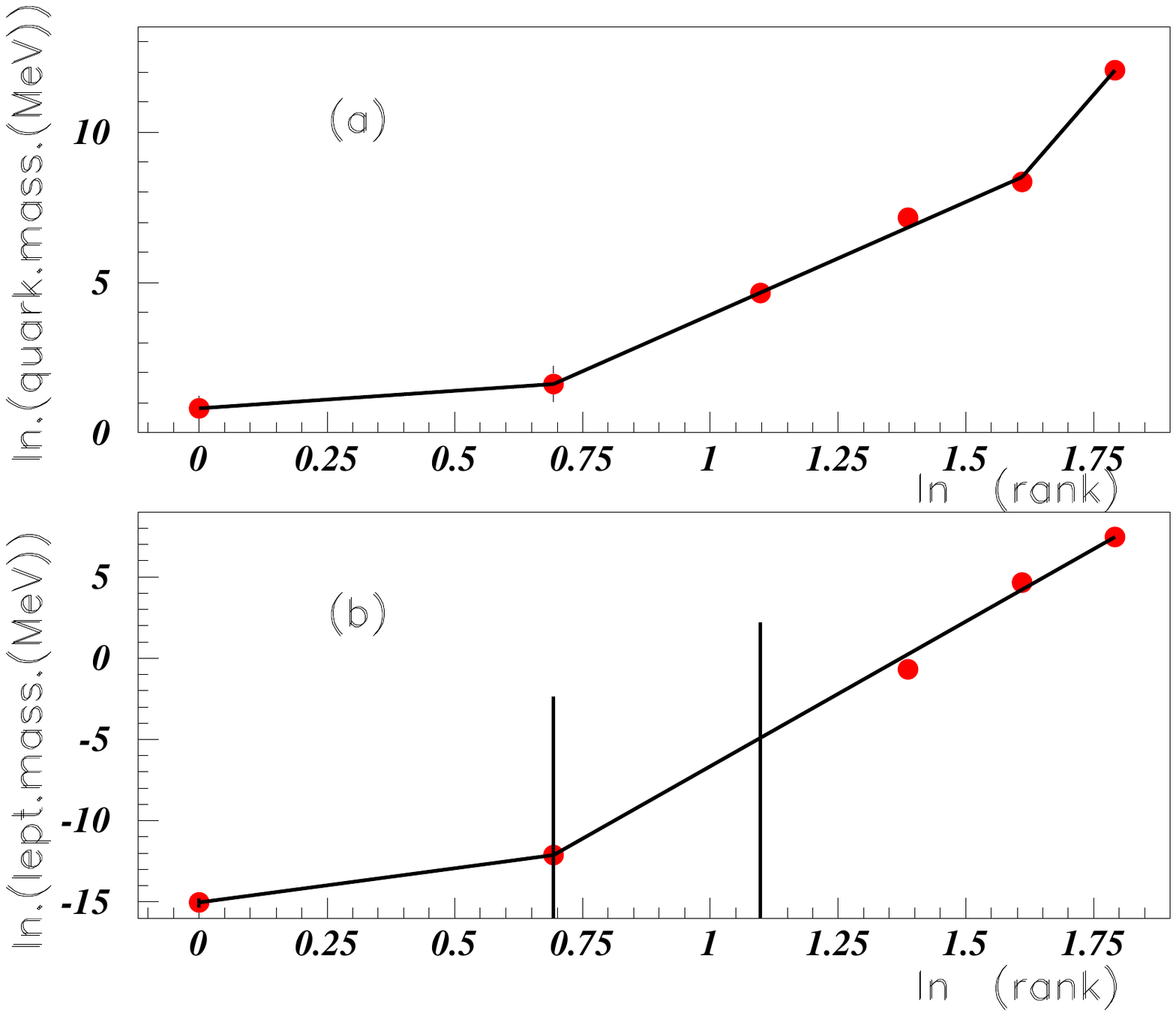}}
\end{center}
\end{figure}
\begin{figure}[ht]
\begin{center}
\caption{Quark over lepton mass ratio}
\hspace*{-3.mm}
\scalebox{0.95}[1]{
\includegraphics[bb=14 240 532 544,clip,scale=0.5]{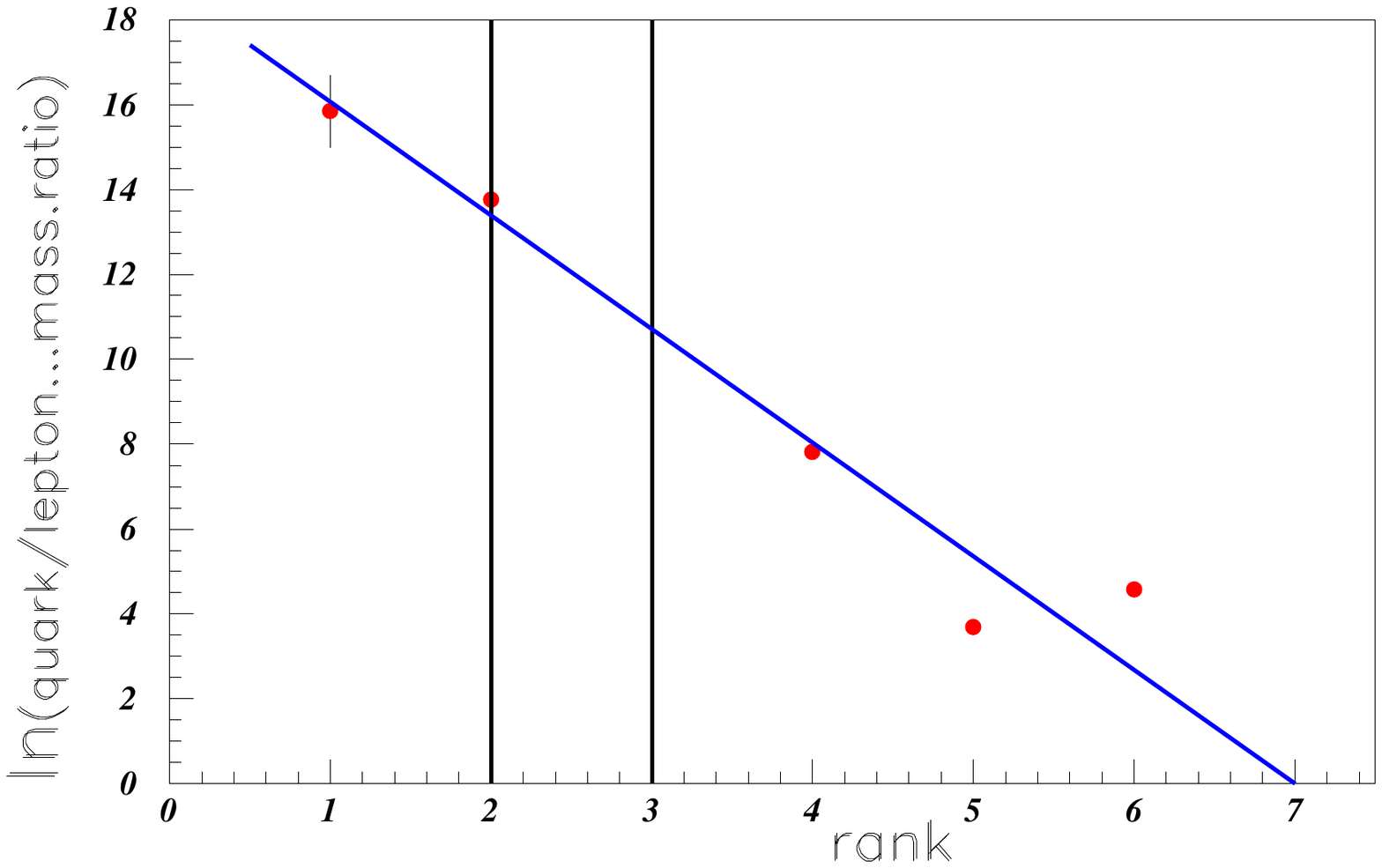}}
\end{center}
\end{figure}
Figure~4 shows the quark over lepton mass ratio, using the mass values mentionned above. This figure predicts the same mass for a hypothetical seventh quark and a hypothetical seventh lepton. This mass equality is observed at best than 5\%, at 
M~$\approx$~270~GeV, by extrapolation of both distributions (inserts (a) and (b)) from figure~5). Notice that extrapolations from figure~3 lead to different hypothetical seventh quark and lepton masses.

 We apply the equation (1) to the elementary particle mass ratio distributions of two adjacent masses:  m$_{r+1}$/m$_{r}$~=~f(r). r is the rank increasing with increased masses. We plot the ratios at 1.5, 2.5 ... for ratios between m(2)/m(1), m(3)/m(2) ...
\begin{figure}[ht]
\begin{center}
\caption{Insert (a) shows the m$_{r+1}$/m$_{r}$ quark mass ratios using the values mentionned before; insert (b) shows the m$_{r+1}$/m$_{r}$ lepton mass ratios using the  values mentionned before
(see text).}
\hspace*{-3.mm}
\scalebox{0.95}[1]{
\includegraphics[bb=10 146 518 558,clip,scale=0.5]{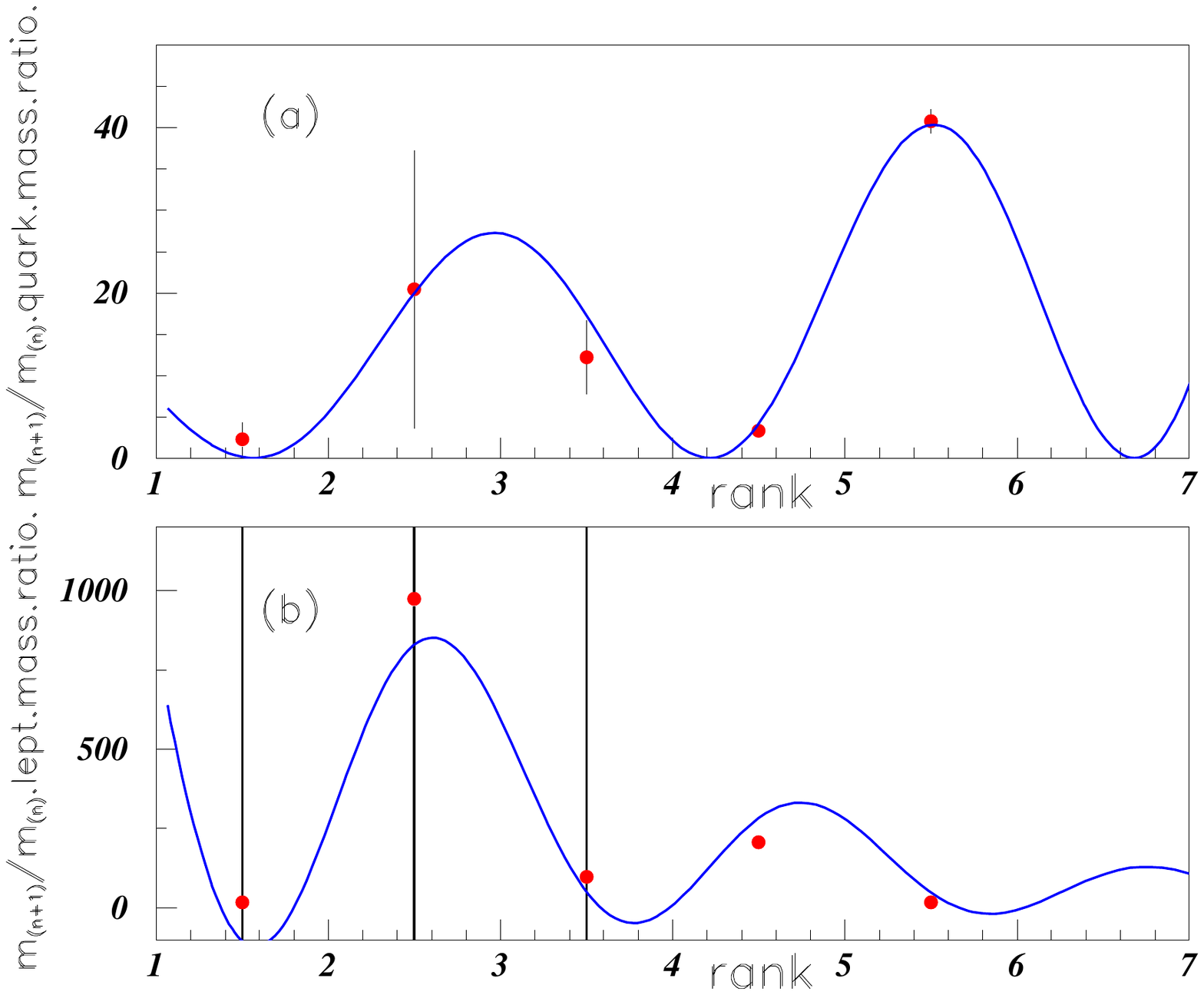}}
\end{center}
\end{figure}
Figure~5 shows the fits for quark (insert (a))  and lepton (insert (b)), of m$_{r+1}$/m$_{r}$ mass ratios using the equation (3). The third neutrino mass, not "calculated", is arbitrarily taken to be 5.2~keV, defined by continuity by comparison with adjacent masses \cite{btib}. The $r_{c}$ undetermination has very few consequences on all parameters, except on $\Omega$. 
 Both distributions are obtained with r$_{c}$~=~40. The same insert (a) is obtained with r$_{c}$~=~30, but then $\Omega$ moves from 14 to 11, therefore $\lambda$ moves from 1.074 up to 1.095. The lepton distribution is obtained with $\Omega$~=~17, $\lambda$~=~1.06. The large error bars in insert (a) are due to the bad knowledge of the "s" quark mass:
104$\pm$30~MeV. The large error bars in insert (b) are the consequence of the bad knowledge of the neutrinos masses. The credibility of the distribution in insert (b) is poor. 

We observe that the distributions shown in figure~5 describe all data points, although  the different slopes in figure~3, indicate different possible $\lambda$ values. 
\section{Conclusion}
In conclusion, we have shown  that fundamental force coupling constants and several atomic energies agree with fractal properties, namely that the log of the function displays a straight line, when plotted versus the log of the rank. 
We observe that in all cases studied here, the first data lies outside the alignment which follows.

 In spite of very low statistics concerning the elementary particle masses (quarks and leptons) and in spite of the fact that the theoretical neutrino masses are quite unknown, we have shown that their masses have also fractal properties. This low statistics involve an undetermination of "r$_{c}$", which is arbitrarily put to 40. The elementary particle masses are well described by discrete scale invariance equation since Re($\alpha)\ll~Im(\alpha)$. The ratios Re($\alpha)$/~Im$(\alpha)$ are equal to 0.062 for quarks and -0.16 for leptons. These data therefore exhibit log-periodic corrections.
\begin{figure}[ht]
\begin{center}
\caption{Log-log plot using gauge boson, quark, and lepton masses (see text).}
\hspace*{-3.mm}
\scalebox{0.95}[1]{
\includegraphics[bb=16 230 523 544,clip,scale=0.5]{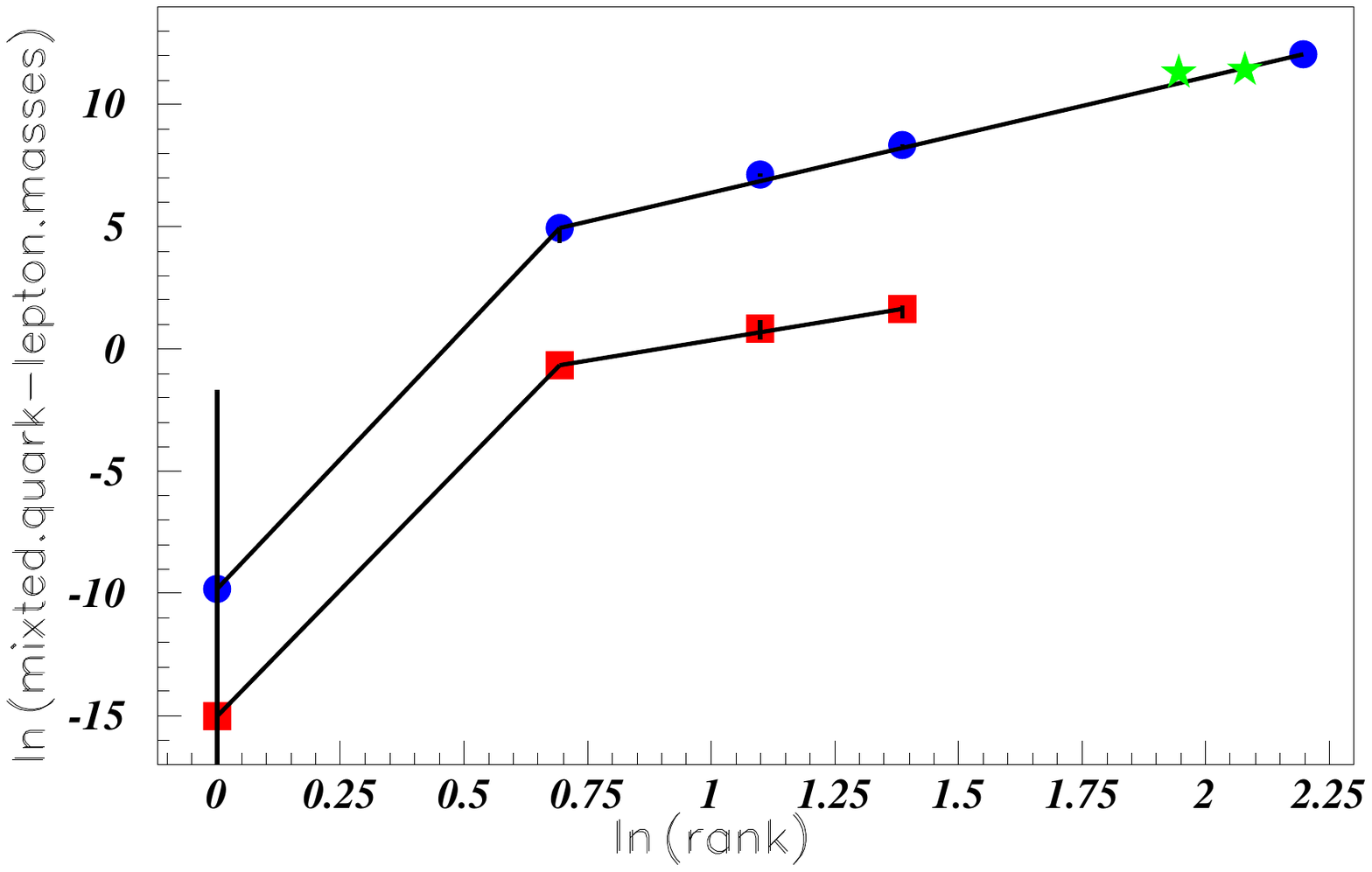}}
\end{center}
\end{figure}
This paper suggests that the quark and lepton masses may not be independant, as shown in figure 4. The same conclusion was found in \cite{btib}. Indeed in \cite{btib}, power laws were presented relying between themselves quark masses, gauge boson masses, and  lepton masses. 
For the quark masses, the fine structure $\alpha$ was used. It was also used together with the proton mass for the gauge boson masses. It was also used for the lepton masses, together with the pion $m_{\pi}$ mass. 
Such relations suggest possible fractal log-log alignements, "mixing" boson, quark, and lepton masses. It was already mentionned that "Quark Lepton Unification (QLU) implies that quark and leptons have the common flavor basis and the common mass basis ..." \cite{hwang}.

Figure~6 shows that is indeed the case.  
The lower alignement with full squares (red on line) is obtained with the following masses: $\nu_{e}$, m$_{e}$, m$_{u}$, and m$_{d}$.  The  upper alignement  with full circles (blue on line) is obtained first with the following masses: $\nu_{\mu}$, m$_{\mu}$, m$_{c}$, and m$_{b}$. Two masses are missing, then the following masses, drawn with full stars (green on line), correspond to the gauge boson W and Z masses;  finally the mass of the top quark is again aligned with the previous ones. The strange quark mass, is not distinguishable from the $\mu$ meson mass. The first two masses of both alignements are proportionnal (parallel lines in log scale)., in a ratio close to 200. The missing masses correspond to rank 5 and 6, and should have respectively M~$\approx$~11650 and 26710~ MeV. They could belong to still unobserved possible additional gauge bosons such as 
those introduced by several theories studying the interaction of standard matter with dark matter \cite{merkel}.

The scale relativity was already applied with success for various applications in physical sciences, astrophysics and others, in living beings, and in human societies \cite{chaline}. The present paper shows that it is also present in various propreties of the sub atomic particles inside quantum physics.
\section{Acknowledgments}
Ivan Brissaud introduces me to the study of hadronic masses using fractal properties. I thank him for his stimulating remarks and interest. Marie-Pierre Comets is acknowledged for critical reading of the manuscript which helps to improve it.
\samepage

\end{document}